\documentclass[sigconf,screen]{acmart}
\AtBeginDocument{%
  \providecommand\BibTeX{{%
    \normalfont B\kern-0.5em{\scshape i\kern-0.25em b}\kern-0.8em\TeX}}}

\acmConference[]{}{}

\usepackage{multirow}
\usepackage{amsmath}
\usepackage{tabularx}
\usepackage{svg}
\usepackage{verbatimbox}
\usepackage[T1]{fontenc}
\usepackage{fancyvrb,listings,caption}
\usepackage[labelformat=simple]{subcaption}
\lstdefinelanguage{kotlin}{
	sensitive=true,
	alsoletter={\%},
	comment=[l]{},
	string=[b]{"},
	keywords=[1]{
fun, class, var
	},
	keywords=[2]{
=, +, return
	},
	keywords=[3]{
Int
	},
}

\lstdefinelanguage{jimple}{
	sensitive=true,
	alsoletter={\%},
	comment=[l]{;},
	string=[b]{"},
	keywords=[1]{
fun, class, var
	},
	keywords=[2]{
=, +, return, public, private
	},
	keywords=[3]{
int, Foo
	},
}

\lstdefinelanguage{goto}{
	sensitive=true,
	alsoletter={\%},
	string=[b]{"},
	keywords=[1]{
	},
	keywords=[2]{
member, @this, @parameter0
},
	keywords=[3]{
int, Foo, signed
	},
}

\usepackage{xcolor}
\usepackage{textcomp}
\definecolor{comment}{RGB}{0,128,0}     
\definecolor{string}{RGB}{255,0,0}      
\definecolor{instruction}{RGB}{0,0,255} 
\definecolor{directive}{RGB}{128,0,128} 
\definecolor{register}{RGB}{128,0,0}    

\lstdefinestyle{nasm}{
	commentstyle=\color{comment},
	stringstyle=\color{string},
	keywordstyle=\color{instruction},
	keywordstyle=[2]\color{directive},
	keywordstyle=[3]\color{register},
	numbers=left,
    frame=trbl,
	numbersep=5pt,
	breaklines=true,
	showstringspaces=false,
	upquote=true,
	tabsize=8,
 xleftmargin=.05\textwidth, xrightmargin=.05\textwidth 
}

\begin{document}

\title{ESBMC-Jimple: Verifying Kotlin Programs via Jimple Intermediate Representation}

\author{Rafael Menezes}
\orcid{0000-0002-6102-4343}
\affiliation{
\institution{University of Manchester}
  \country{United Kingdom}
}
\additionalaffiliation{
  \institution{Federal University of Amazonas}
  \country{Brazil}
}

\author{Daniel Moura}
\orcid{0000-0002-8116-268X}
\affiliation{%
  \institution{Federal University of Amazonas}
  \country{Brazil}
}

\author{Helena Cavalcante}
\orcid{0000-0001-9632-3637}
\affiliation{%
 \institution{Federal University of Amazonas}
 \country{Brazil}
 }

\author{Rosiane de Freitas}
\orcid{0000-0002-7608-2052}
\affiliation{%
  \institution{Federal University of Amazonas}
  \country{Brazil}}

\author{Lucas C. Cordeiro}
\orcid{0000-0002-6235-4272}
\affiliation{
  \institution{University of Manchester}
  \country{United Kingdom}
}
\additionalaffiliation{
  \institution{Federal University of Amazonas}
  \country{Brazil}
}


\begin{abstract}
We describe and evaluate the first model checker for verifying Kotlin programs through the Jimple intermediate representation. The verifier, named ESBMC-Jimple, is built on top of the Efficient SMT-based Context-Bounded Model Checker (ESBMC). It uses the Soot framework to obtain the Jimple IR, representing a simplified version of the Kotlin source code, containing a maximum of three operands per instruction. ESBMC-Jimple processes Kotlin source code together with a model of the standard Kotlin libraries and checks a set of safety properties. Experimental results show that ESBMC-Jimple can correctly verify a set of Kotlin benchmarks from the literature; it is competitive with state-of-the-art Java bytecode verifiers. A demonstration is available at \url{https://youtu.be/J6WhNfXvJNc}.
\end{abstract}


\maketitle

\begin{figure*}[!ht]
  \includegraphics[width=0.9\textwidth]{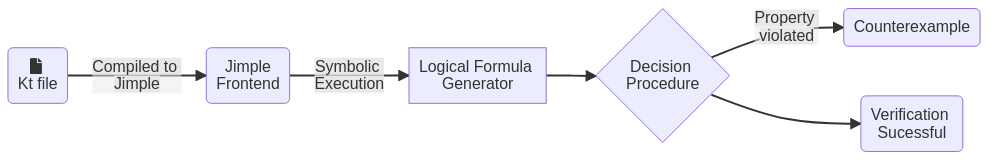}
  \caption{Architectural overview of ESBMC with its extension (\textbf{Jimple Frontend}) for verifying Kotlin programs.}
  \Description{}
  \label{fig:esbmc_jimple_overview}
\end{figure*}

\section{Introduction}

Kotlin is a multiplatform programming language~\cite{kotlin} to provide a more productive way to develop software on top of the JVM~\cite{gois2019empirical}. Kotlin adds functional programming features and safety checks; it has full interoperability with JVM~\cite{kotlin}. Google added support for Android Development using Kotlin, becoming one of the popular choices among software developers~\cite{9054859}. However, as the Kotlin programming language's popularity grows, there is a clear need for more associated verification tools to ensure its safety.



Bounded Model Checking (BMC) is a formal verification technique that can check implementation errors in software~\cite{clarke2018handbook}. The basic idea of the BMC technique is to check (the negation of) a given property at a given depth. A significant strength of BMC is that it analyzes only bounded program runs, thereby achieving decidability. A notable example of a tool that implements BMC is ESBMC~\cite{ESBMC}, which is based on Satisfiability Modulo Theories (SMT) solvers and checks safety properties such as arithmetic overflow, array bounds, pointer safety, and user-specified properties. ESBMC uses its intermediate representation (IR) GOTO to analyze programs, converting loop structures into the GOTO form and supporting object-oriented programming (OOP) features.


We rely on the Jimple IR, a stackless 3-address IR with a set of instructions from the Soot framework~\cite{vallee2010soot} to simplify the bytecode analysis. As a result, ESBMC does not track a virtual table, and neither maintains a stack~\cite{arlt2013joogie}, making the verification process more straightforward. Since ESBMC verifies C/C++~\cite{ESBMC,monteiro2022model} programs, a similar approach can be used when dealing with Jimple, as the symbolic execution engine has the support for OOP features. Some verifiers support Java bytecode verification (e.g., JBMC~\cite{cordeiro2018jbmc} and JayHorn~\cite{KahsaiRS19}). However, they have shortcomings when verifying Kotlin because of limitations on absent models and assertions, e.g., array initialization on Kotlin relies on its SDK for the implementation.

We describe the first model checker for Kotlin programs, ESBMC-Jimple, which adds Jimple support to ESBMC to verify safety properties in Kotlin programs. Therefore, we can use ESBMC's verification strategies (e.g., incremental BMC and \textit{k}-induction) and optimizations (e.g., program slicing) to enable the verification of Kotlin programs that otherwise would exhaust the systems resources. To evaluate  ESBMC-Jimple, we developed a new set of benchmarks for Kotlin analysis extracted from a GitHub repository~\cite{githubRepo} and also by translating a subset of the SV-COMP benchmarks from Java safety into Kotlin~\cite{beyer2022progress}. Our experimental evaluation has shown that ESBMC-Jimple could correctly verify more Kotlin benchmarks than JBMC~\cite{cordeiro2018jbmc}, a native Java bytecode verifier.

\section{Tool Description}
\label{tool-description}

\subsection{Architecture and Implementation}

Fig.~\ref{fig:esbmc_jimple_overview} illustrates the ESBMC-Jimple architecture. The flow starts from a Kotlin file, compiled to Jimple (through Soot), and then symbolically executed to produce a logical formula. 

\subsection{Jimple Convertion Methodology}

To use the ESBMC symbolic execution, we convert the original Jimple IR into GOTO, which is the IR used by ESBMC. It contains instructions for function calls, assignments, declarations, exception handling, and thread creation. For the Jimple IR support, a subset of this language was used (Table~\ref{table:goto-subset}). Further instructions are not used yet (e.g., try/catch blocks), and some instructions are used internally for analysis (e.g., \textit{assume} and \textit{assert} statements). Listing~\ref{lst:goto-grammar} shows how a GOTO program is represented through named \textit{Functions}, which are implemented using \textit{Statements} and \textit{Expressions}.
\begin{lstlisting}[
 caption=GOTO grammar subset.\label{lst:goto-grammar}]
GOTO       := Functions* 
Functions  := Statements* END_FUNCTION
Statements := Var=Expr | assert Expression 
            | goto Label | ... 
Expression := Var | Const | Var Expression 
            | Expression + Expression | ...
\end{lstlisting}

\begin{table}[htb]
	\caption{\label{table:goto-subset} GOTO subset supported by ESBMC-Jimple.}
			\begin{tabularx}{0.45\textwidth}{|X|X|}
				\hline
				\textbf{Statement} &
				\textbf{Description} \\
				\hline
				ASSIGN($a$, $b$) & Assigns value of $b$ into symbol $a$. \\
				\hline
				DECL($a$, $t$) & Create a new valid memory object for $a$ of type $t$. \\
				\hline
				DEAD($a$) & Deletes the $a$ memory object, making it an invalid-object. \\
				\hline
				FUNCTION\_CALL($n$, $\ldots$) & Calls function $n$ with parameters. \\
				\hline
				LABEL($L$) & Sets a label $L$ in the code, this label can be used for jumps. \\
				\hline
				GOTO($L$) & Goto a specific label $L$. \\
				\hline
				IF($e$, $L$) & Checks if a condition $e$ is true, and if it is goto label $L$.  \\
				\hline
				SKIP & Go to the next instruction.  \\
				\hline
				RETURN($e$?) & Return to the latest FUNCTOIN\_CALL with the optional result $e$.  \\
				\hline
				THROW($e$) & Throws an exception with expression $e$.  \\
				\hline
			\end{tabularx}
\end{table}

The GOTO language uses \textit{Expressions}. Those expressions range from binary operations (e.g., add and subtraction) to dereferencing operations (e.g., accessing a member of the struct). Since there is a C frontend, the most needed expressions are implemented. The exceptions are the \texttt{cmp} operations that exist for Java-based languages, implemented during the GOTO conversion as an expression (Expr).

\subsubsection{Objects and Classes}
\label{sub:objectsandclasses}

Jimple contains OOP elements; every file is defined via class members, which include fields and functions that can be \textit{virtual} or \textit{static}. These members can be overridden during inheritance; functions with the same name can have multiple specializations by changing their parameters. To model classes, we rely on a similar approach to verify C++ programs~\cite{monteiro2022model}. We map the virtual fields into structs and the static fields into global variables (within a scoped namespace). However, the GOTO language does not support polymorphism and neither for function inside a structure. To solve this, we used the following approach. (1) Generate an identifier for each function using its name and parameter types. A function \texttt{foo} with return type
\texttt{int} and parameter type of \texttt{double} is renamed to \texttt{foo\_int\_double}. (2) Due to how Jimple handles functions parameters, we convert each parameter as a global variable; at the start of each function this value is then read. This operation is marked as atomic for the GOTO flow. (3) For virtual functions, we add an extra argument into the function: a pointer of its own class. A virtual function \texttt{bar()} inside the class \texttt{A} would become \texttt{bar(A*)}. This parameter behave as the \texttt{this} pointer of the object.

This approach simplifies the inheritance process as the inherited classes have the same fields that can be used for access. Jimple IR handles the virtual table, stating from which class of the inheritance tree the method should be called. For the object memory model, we use the one for C++ programs~\cite{ESBMC}. In this model, primitive types are allocated into the \textit{stack}; constructed objects (i.e., \textit{new} or \textit{newarray}) use the \textit{heap} concept. We do not model a garbage collector; we assume that it will be valid for the entire verification flow once something is allocated into the heap.

\subsubsection{Jimple Convertion}

The Jimple conversion begins with generating types for the
classes; every class is converted into a struct containing all its virtual fields. Each Jimple statement is converted into an equivalent statement in GOTO for functions. Table~\ref{table:jimple-translation} contains the translations used for every Jimple statement; it makes assumptions of previously conversions performed: \emph{DeclarationLists} (e.g., \texttt{int a,b,c;}) were converted into a multiple individual declarations, \emph{at identifiers} (e.g. \texttt{@this}, \texttt{@param1}) were injected properly.

\begin{table}[]
	\caption{\label{table:jimple-translation} Jimple statements translation in ESBMC-Jimple.}
			\begin{tabularx}{0.45\textwidth}{|X|X|}
				\hline
				\textbf{Jimple} &
				\textbf{GOTO} \\
				\hline
				declaration($a$, $t$) & DECL($a$,$t$) \\
				\hline
				label($L$) & LABEL($L$) \\
				\hline
				breakpoint & SKIP \\
				\hline
				virtualinvoke(\textit{object}, \textit{parameters} & FUNCTION\_CALL(\textit{object}, \textit{parameters})  \\
				\hline
				specialinvoke(\textit{object}, \textit{parameters} & FUNCTION\_CALL(\textit{object}, \textit{parameters}) \\ 
				\hline
				staticinvoke(\textit{parameters}) & FUNCTION\_CALL( \textit{parameters} ) \\ 
				\hline
				return($e$?) & RETURN($e$?) \\ 
				\hline
				$v=e$ & ASSIGN($v$, $e$) \\ 
				\hline
				if($e$) goto $L$  & IF($e$, $L$) \\ 
				\hline
			    throw($e$) & THROW($e$)\\ 
				\hline
			\end{tabularx}
\end{table}

\subsubsection{Illustrative Example}

We use the example in Listing~\ref{lst:examplekt} to show how the conversion process works. There exists a class named \texttt{Foo}, which contains a mutable field \texttt{member} and the \texttt{increment} method, which increments the \texttt{member} field. After the Jimple file is generated via Soot~\cite{vallee2010soot}, ESBMC can verify it by invoking:
\begin{center}
    \texttt{esbmc <file>.jimple -{}-k-induction -{}-overflow-check}
\end{center}

\noindent where \texttt{<file>.jimple} is the Jimple file to verify, \texttt{-{}-overflow-check} enables checking for signed integer overflows, and \texttt{-{}-k-induction} sets the \textit{k}-induction as the proof rule. The full list of options can be seen by using the flag  \texttt{-{}-help}.

\begin{lstlisting}[
 caption=Example of a Kotlin program that defines the Class \texttt{Foo} used to illustrate the GOTO conversion.\label{lst:examplekt},
 language=kotlin, style=nasm]
class Foo(var member: Int) {
  fun increment(N: Int): Int {
    member = member + N
    return member }}
\end{lstlisting}
Listing~\ref{lst:examplegoto} contains the GOTO version of the \texttt{increment} function in Listing~\ref{lst:examplekt}. Here, we see the function name (as described in Subsection~\ref{sub:objectsandclasses}). Next, Jimple adds intermediate variables (e.g., $r0$, $i0$, $i1$). The ``at'' identifiers, i.e., ``@this'' and ``@parameter0'' indicate the object pointer to itself and the first function parameter, respectively.

\begin{lstlisting}[caption=Increment function from converted program. The function is the result of the GOTO conversion of the \texttt{increment} method of the Listing~\ref{lst:examplekt}.\label{lst:examplegoto},language=goto, style=nasm]
increment_2 (Foo:increment_2):
    Foo* r0; 
    signed int i0, $i1, $i2, $i3;
    r0=(Foo*)@this; i0=@parameter0; 
    $i1=r0->member;
    $i2=$i1 + i0; r0->member=$i2;
    $i3=r0->member; RETURN: $i3
\end{lstlisting}

ESBMC will unroll the program during the symbolic execution, converting it to an SSA form. 
In Listing~\ref{lst:examplekt}, we can check the \texttt{increment} function; we make its arguments non-deterministic. The verification conditions are then encoded in the form of $C \wedge \neg P$, as shown in Listing~\ref{lst:vcc}:
\begin{lstlisting}[
 caption=Verification condition.\label{lst:vcc}, escapeinside={(*}{*)}]
C := @this=nondet()
     (*$\wedge$*) r0=@this (*$\wedge$*) @parameter0=nondet()
     (*$\wedge$*) i0=@parameter0 (*$\wedge$*) $i1=r0->member 
P := overflow("+", i0, $i1)
\end{lstlisting}

$C$ contains the assignments; both function arguments, i.e., \texttt{@this} and \texttt{@parameter0} are set as nondeterministic. $P$ is the property that adding $i0$ with $\$i1$ can not lead to an overflow. 

\section{Experimental Evaluation}
\label{experiments}


We evaluated ESBMC-Jimple using a benchmark suite and compared its performance against JBMC~\cite{cordeiro2018jbmc} since it supports Java Bytecode verification efficiently. Other tools were considered, Jayhorn~\cite{KahsaiRS19} and Joogie~\cite{arlt2013joogie}. However, we could not use JayHorn and neither Joogie for our benchmarks. 
Tools, benchmarks, and evaluation results are available on a supplementary web page.\footnote{\url{https://doi.org/10.5281/zenodo.6514235}}.




\textit{Description of Benchmarks.} We developed a small suite of benchmarks (see Table~\ref{table:Result}) for evaluating ESBMC-Jimple. These benchmarks contain execution paths that can trigger bugs in Kotlin applications and execution paths without property violations. They also include the following properties: reachability, overflows, and null-pointer exception. The benchmarks contain non-deterministic behavior, modeled through the Java \textit{Random} function.


\subsection{Objective and Setup}


Our main experimental questions are:
\begin{enumerate}
\item[\textbf{EQ1}]: (\textbf{soundness}) Can the tool prove (partial) correctness?
\item[\textbf{EQ2}]: (\textbf{performance}) How long does ESBMC-Jimple take to verify a Koltin application? 
\item[\textbf{EQ3}]: (\textbf{completeness}) Can the tool correctly identify bugs in Kotlin programs?
\end{enumerate}

We set up the experiments on a Ubuntu 20.04 with $160$ GB of RAM running a 25-core Intel KVM CPU. If the tool could produce a counterexample, it is manually tested on the benchmark. Importantly, all presented execution times are CPU times, i.e., only the elapsed periods spent in the allocated CPUs, measured with the times system call (POSIX system). Since neither ESBMC-Jimple nor JBMC supports Kotlin input directly, we first compile the Kotlin source file into Java Bytecode and then convert it back to Jimple. Then, the Java Bytecode is given as input to JBMC; the Jimple is given for the ESBMC-Jimple.



\subsection{Results}

Table~\ref{table:Result} shows the results; ESBMC-Jimple could verify all benchmarks originating from Kotlin programs correctly. Note that JBMC was not developed for verifying Kotlin programs, which thus did not handle some constructs correctly. For example, regarding the overflow benchmarks (TC0 and TC1), ESBMC-Jimple could detect the overflow in both cases; nondeterministic values were used for the addition of positive numbers (i.e., $X + Y \geq 0, \forall X \geq 0,Y \geq 0$). ESBMC-Jimple could produce a counterexample that led to the overflow. For TC2 and TC3, we use the same approach, but for negative numbers (i.e., $X + Y \leq 0, \forall X \leq 0,Y \leq 0$), ESBMC-Jimple could also produce counterexamples. However, JBMC was unable to identify any violations in TC0-TC3. 

Division by zero benchmarks (TC4 and TC5) ensure that the denominator of a division could never be zero. ESBMC-Jimple and JBMC could correctly identify the flaws and produce the respective counterexamples. Similarly, for array bounds check benchmarks (TC6-TC9), ESBMC-Jimple and JBMC had the same outcome; they could detect bounds violations for nondeterministic arrays originating from Kotlin programs. We also evaluated benchmarks with user assertion violations (T10-TC15) and without violation (TC16-TC20). ESBMC-Jimple could verify these benchmarks correctly, but JBMC refuted the assertions, incorrectly triggering a violation for all benchmarks stated as safe.

Both EQ1 and EQ3 can be confirmed through our experiments, as ESBMC-Jimple could reason over the safety of Kotlin programs and generate valid counterexamples that trigger the property violation. Regarding performance (EQ2), due to ESBMC's efficient support for bug finding and proof strategies (e.g., incremental BMC, \textit{k}-induction), ESBMC-Jimple could quickly refute safety properties or prove (partial) correctness of Kotlin programs.

\begin{table}[!ht]
\caption{Experimental results, where column ``Found'' indicates whether a bug was detected, followed by a column ``CE'' showing whether a counterexample was provided. The bottom lines are the results summary. 
}
\begin{center} 
\begin{tabular}{|c|c|c|c|c|c|c|c|c|c|c|c|}
\hline 
\multicolumn{2}{|c|}{\textbf{Benchmark}} &\multicolumn{2}{c|}{\textbf{JBMC}} &\multicolumn{2}{c|}{\textbf{ESBMC-Jimple}} \\ \hline
       \textbf{IDs} & \textbf{Property}             &\textbf{Found}  &\textbf{CE}   &\textbf{Found}  &\textbf{CE} \\\hline
        TC0-1  & Overflow  &No     &No    &Yes     &Yes  \\\hline
        TC2-3  & Underflow       &No     &No   &Yes     &Yes   \\\hline
        TC4-5  & Div-by-zero       &Yes     &Yes   &Yes     &Yes  \\\hline
        TC6-9  & Out-of-bounds       &Yes    &Yes &Yes     &Yes \\\hline
        TC10-15 & Assertion Fail        &Yes    &No &Yes     &Yes \\\hline
        TC16-20 & No violation         &Yes    &N/A  &No    &N/A \\\hline
        \multicolumn{2}{|c|}{\textbf{Correct Results}}  &\multicolumn{2}{c|}{$57\%$} &\multicolumn{2}{c|}{$100\%$} \\\hline
        \multicolumn{2}{|c|}{\textbf{Confirmed Results}} &\multicolumn{2}{c|}{$38\%$} &\multicolumn{2}{c|}{$100\%$} \\\hline
        \multicolumn{2}{|c|}{\textbf{Total CPU Time}} &\multicolumn{2}{c|}{$3.563s$} &\multicolumn{2}{c|}{$18.878s$} \\\hline

\end{tabular}
\label{table:Result}
\end{center}
\end{table}

\subsection{Threats to Validity}

Compilers and decompilers might introduce (or remove) bugs during the translation. Our approach has three translations: (1) Kotlin programs compiled into JVM; (2) Soot decompiling them into Jimple; and (3) ESBMC translating Jimple onto GOTO. Any of those phases can change the program's semantic behavior. Additionally, ESBMC-Jimple relies on operational models (OMs) to verify a program; we developed the OM for a subset of Kotlin and Java standard libraries. This subset was chosen based on features needed on the evaluated benchmarks. Those OMs might approximate the original program's behavior, leading to an invalid program encoding. Lastly, ESBMC-Jimple relies on the same memory model used for ESBMC's C/C++ analysis. This model can limit the behavior (e.g., garbage collection) that a program can have.

\section{Related Work}
\label{related-work}

Some tools also leverage Java verification using bytecode for the analysis, mainly JBMC~\cite{cordeiro2018jbmc} and JayHorn~\cite{KahsaiRS19}. JBMC has an architecture similar to ESBMC-Jimple, having to translate Java Bytecode into a GOTO program. JayHorn is a software verifier for Java bytecode that generates a set of constrained Horn clauses. The architecture of JayHorn contains the Soot framework as its front-end, with transformation validated with Randoop. It then uses a CHC solver to check the input program safety. Unfortunately, both JBMC and JayHorn have shortcomings (as seen in our experiments) when verifying Kotlin code. Another similar work is Joogie~\cite{arlt2013joogie}, which translates Jimple code into the Boogie language. Boogie is a description language that can be translated into SMT formulas to be checked by an SMT solver. Joogie focuses on verifying Java applications and could verify real-world Java programs successfully.



\section{Conclusions and Future Work}
\label{conclusions}

We presented and evaluated ESBMC-Jimple, the first software model checker for Kotlin applications, which relies on the Jimple IR from the compiled Kotlin program and the ESBMC verification engine. ESBMC-Jimple can handle various features from Kotlin, including classes, inheritance, and polymorphism. Furthermore, ESBMC-Jimple outperformed another state-of-the-art Java bytecode verifier since it could detect and generate more real counterexamples for the evaluated benchmarks. We will support exception handling, more operational models for the Kotlin standard, and extend the range of properties for future work.

\begin{acks}
This research was partially sponsored by Motorola Mobility Comércio de Produtos Eletrônicos Ltda and Flextronics da Amazônia Ltda, according to Federal Law nº 8.387/1991, through agreement nº 004/2021, signed with ICOMP/UFAM.
\end{acks}

\bibliographystyle{ACM-Reference-Format}
\bibliography{sample-base.bib}
\end{document}